\documentclass[pre,nofootinbib,twocolumn,floatfix,showpacs,sort&compress,superscriptaddress]{revtex4}
\usepackage{graphicx}
\usepackage{amsmath}
\usepackage{amssymb}

\ifnum\lefthyphenmin<2\lefthyphenmin=2\fi
\ifnum\righthyphenmin<2\righthyphenmin=2\fi

\begin{document}

\title{Denaturation of Circular DNA: Supercoil Mechanism}
\author{Amir Bar}
\affiliation{Department of Physics of Complex Systems, The Weizmann Institute of Science, Rehovot 76100, Israel}
\affiliation{Department of Computer Science and Applied Mathematics,The Weizmann Institute of Science, Rehovot 76100, Israel}
\author{Alkan Kabak\c{c}{\i}o\u{g}lu}
\affiliation{Department of Physics, Ko\c c University, Sar\i yer 34450 \. Istanbul, Turkey}
\author{David Mukamel}
\affiliation{Department of Physics of Complex Systems, The Weizmann Institute of Science, Rehovot 76100, Israel}

\begin{abstract}
The denaturation transition which takes place in circular DNA is
analyzed by extending the Poland-Scheraga model to include the
winding degrees of freedom. We consider the case of a homopolymer
whereby the winding number of the double stranded helix, released by
a loop denaturation, is absorbed by \emph{supercoils}. We find that
as in the case of linear DNA, the order of the transition is
determined by the loop exponent $c$. However the first order
transition displayed by the PS model for $c>2$ in linear DNA is
replaced by a continuous transition with arbitrarily high order as
$c$ approaches 2, while the second-order transition found in the
linear case  in the regime $1<c\le2$ disappears. In addition, our
analysis reveals that melting under fixed linking number is a
\emph{condensation transition}, where the condensate is a
macroscopic loop which appears above the critical temperature.
\end{abstract}

\date{\today}


\maketitle

\section{Introduction}

Thermal denaturation of DNA is a process by which the two strands of
the molecule unbind upon heating. A good understanding of the
underlying physics is relevant to certain biological systems (e.g.,
thermophilic organisms \cite{DSD1992,HS2004}) as well as synthetic
technologies ~\cite{WB1985} such as polymerase chain reaction (PCR)
\cite{HM1994,HM1996} and DNA microarrays \cite{CH2006}.  The unbinding
transition takes place at a specific temperature, coined melting or
denaturation temperature, which can be defined experimentally as the
temperature at which the fraction of unbound base pairs reaches, say,
half of its maximal value. For a relatively homogenous DNA chain
composed largely of A-T (or G-C) pairs, melting takes place through a
very sharp increase in the fraction of broken bases, suggesting a
first-order phase transition in an idealized homogeneous system. This
phase transition has been investigated by means of
  various theoretical approaches developed in recent
  decades~\cite{PB1989,Fisher1966,PS1966,DPB1993,KMP2000, WM1972, Ben1980, Ben1992,
     BM2000, PMD2007, MS1994b, NG2006}.

A prototypical model employed in theoretical studies of this
phenomenon is the Poland-Scheraga (PS) model~\cite{PS1966} in which a
microscopic configuration of the DNA molecule is described by an
alternating succession of bound segments (dsDNA) and denaturated loops
(ssDNA). As the temperature is increased the total length of the bound
segments decreases, eventually vanishing at the melting
transition. The transition is a result of the competition between the
enthalpy associated with the hydrogen bonding of the matching bases,
and the entropy of loops. The loop entropy has the asymptotic form
$\sim s^{l}/l^{c}$ for large loop size $l$, where $s$ is a geometric,
non-universal constant and $c$ is a universal exponent.  The original
PS model makes the simplifying assumption that the binding energy is
the same for all base pairs, in which case the nature of the
transition depends only on the parameter $c$~\cite{PS1966}. For
$c\le1$ no transition takes place and the two strands are bound at all
temperatures. For $1<c\le2$ the model exhibits a second-order
  melting transition where the average loop length increases and
  becomes macroscopic of order $L$ as the critical point is approached from below.  For
  $c>2$ the transition is first order and the average loop length
  remains $O(1)$ for $T\le T_c$. For $T>T_c$ a macroscopic loop, formed abruptly at
  $T_c$, is present. In $d=3$ dimensions and with exclusion interaction properly
taken into account, one obtains $c \approx 2.12$
\cite{KMP2000,KMP2002,COS2002} and the transition is predicted to be
first order. The PS model has later been extended to address the
sequence dependence of the melting transition in heteropolymeric
DNAs~\cite{Meltsim}.

The DNA molecule is helical, and therefore denaturation entails
unwinding of the two strands around one another. The PS model
ignores this fact, as the elastic strain can be relaxed by the
rotation of the chain ends. However there are cases where the
helicity can not be ignored. For example, bacteria have circular
DNAs (plasmids) whose \emph{linking
  number} (the number of times one strand winds around the other) is a
topological invariant. Similarly, certain single-molecule
experiments require the chain ends to be rotationally constrained.
In such cases, unwinding of a loop is possible only if some extra
linking number can be absorbed by the rest of the molecule.

Previous studies that model denaturation of circular DNA proposed two
mechanisms by which bound DNA segments may host extra linking
number released by opening a loop: (a) increasing the \emph{twist }(the excess stacking angle
integrated along the centerline) \cite{RB2002, GOY2004};
or (b) increasing the \emph{writhe }(which is a function of the
centerline configuration itself), for example by forming a
{\it supercoil} \cite{KOM09,KOM2010,KAS2010}. The AFM images of thermally denatured DNA circles
adsorbed on a mica surface suggest that supercoils do form in
conjunction with denaturation loops~\cite{YI2002}. Numerical
studies similarly point at the writhe as the dominant mechanism for
absorbing the extra linking number in long DNA circles~\cite{KAS2010}.

In this paper we study in detail the case of supercoils. In an earlier
work this model has been studied at temperatures below the melting
point \cite{KOM09}, by means of a grand canonical treatment where the
{\em expectation value} of the linking number is fixed. Here we
generalize this approach and further consider the high-temperature
denatured phase in order to study the nature of the melting
transition. The validity of our results is then verified by a direct
calculation within a canonical formalism where the linking number is
strictly conserved. This approach allows us to point out an
inconsistency in the assumed analogy with the PS model in
Ref.\cite{KOM09}. Finally, we find the following phase diagram: For
$c\le2$ the model exhibits no phase transition and a steady increase
of loop fraction with temperature. For $c>2$ a continuous transition
of order $\left\lceil\frac{c-1}{c-2}\right\rceil$ takes place, where
$\left\lceil q \right\rceil$ is the upper integer value of $q$. The
order of the transition tends to infinity as $c\to2$.

The paper is arranged as follows: In section \ref{sec:modeldef}, we present the
model. In section \ref{sec:gcantreat}, the denaturation transition is first established
in the grand-canonical ensemble, where we introduce a
regularization scheme used earlier in \cite{KMP2002}. This procedure
allows us to draw an analogy between the high-temperature phase and a
Bose-Einstein condensate where a critical fluid (microscopic loops)
coexists with a condensate (a single macroscopic loop). In section \ref{sec:canonical},
we reinvestigate the model within the canonical formalism: while we
observe a general agreement between the two ensembles, we also point
out a difference between the corresponding condensates that suggests
the inequivalence of the two ensembles for finite systems in the present
context. Finally, in section \ref{sec:conclusions}, we present some concluding remarks and
discuss possible future directions.

\section{Model definition}
\label{sec:modeldef}
\begin{figure}
\begin{centering}
\includegraphics[scale=0.6]{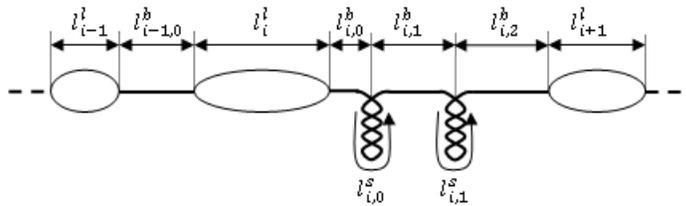}
\par\end{centering}

\centering{}\caption{\label{fig:configuration} A typical configuration
  of a portion of the circular DNA model used in this study.
}

\end{figure}

Following \cite{KOM09} we extend the PS model to
include supercoiled DNA segments. Thus, a microscopic configuration is
composed of an alternating arrangement of three types of segments:
\begin{enumerate}
\item a \emph{bound segment}, in which base pairs are intact but no
  supercoiling takes place.
Following the PS model, we neglect the entropic contribution of such a
segment, so that its Boltzmann weight is solely determined by the
binding energy $E_{b}<0$ and the segment length $l$ as $e^{-\beta l
  E_{b}}\equiv\omega^{l}$, where $\beta=1/k_{B}T$.
\item a \emph{loop}, in which pairing is sacrificed in favor of entropy
  as the persistence length of ssDNA is roughly 10 times shorter than
  that of dsDNA. The associated Boltzmann weight is purely entropic and
  asymptotically given as
  $\Omega(l)=A\frac{s^{l}}{l^{c}}$, where $s$ is a
  geometrical factor and $A$ is a constant coined the ``cooperativity
  parameter'' ~\cite{PS1966}. The (universal) \emph{loop exponent } $c \approx 2.12$ is
  determined by the dimensionality of the embedding space ($=3$) and
  the connective topology of the polymer system~\cite{KMP2000,Dup1986}.

\item a \emph{supercoil}, in which two halves of a dsDNA segment wind
  around each other (see Fig.\ref{fig:configuration}). The
  corresponding Boltzmann weight is given by $e^{-l\beta
    E_{s}}\equiv\nu^{l}$, where $E_s$ ($0>E_s>E_b$) is the energy gain
  of a base pair in a supercoiled segment. Our model reduces to
    the PS model when $\nu=0$.
\end{enumerate}
It is assumed that supercoils occur within bound regions only and
hence a loop is always terminated by two bound segments (of type 1
above). A typical configuration of part of a circular DNA molecule is
shown in Fig.\ref{fig:configuration} where $l_{i}^{l}$ denotes the
length of the $i^{th}$ loop, while $l_{i,j}^{b}$ and $l_{i,j}^{s}$
stand for the lengths of the $j^{th}$ bound segment and the $j^{th}$
supercoil following the $i^{th}$ loop, respectively. The Boltzmann
weight corresponding to the configuration in Fig.\ref{fig:configuration} is

\begin{eqnarray*}
\Omega\left(l_{i-1}^{l}\right)\times\omega^{l_{i-1,0}^{b}}\times\Omega\left(l_{i}^{l}\right)\times\omega^{l_{i,0}^{b}}\times\nu^{l_{i,0}^{s}}\times\\ \omega^{l_{i,1}^{b}}\times\nu^{l_{i,1}^{s}}\times\omega^{l_{i,2}^{b}}\times\Omega\left(l_{i+1}^{l}\right)\end{eqnarray*}

Let $L_{b},L_{s}$ and $L_{l}$ be the {\em total} length of bound,
supercoil and loop segments respectively. The length of the DNA is
given by $L_{b}+L_{s}+L_{l}=L$. The conservation of the linking number
is imposed by the additional condition that an increase in the total
loop length $L_l$ (reducing the linking number) is compensated by a
proportional increase in the total supercoil length $L_s$ (recovering
the linking number), and vice versa. Given the ground state $L_b=L$,
$L_s = L_l = 0$, this yields the constraint $L_{s}=\alpha
L_{l}$. $\alpha$ is the proportionality constant and for simplicity we
assume here $\alpha=1$, though the result is qualitatively the same
for other values \cite{KOM2010}. In this model, as in the PS case, it
is more convenient to work within a grand canonical ensemble, where
the above constaint is relaxed to the equality of corresponding
ensemble averages, i.e., $\langle L_s \rangle = \langle L_l \rangle $.

\section{Grand Canonical Treatment}
\label{sec:gcantreat}

For completeness we first outline the derivation in \cite{KOM09} for
this case. To account for the two constraints above,
the grand partition sum is constructed as a function of two fugacities $z$ and $\mu$ as
\begin{equation}
Q(z,\mu)=\sum_{L_{b},L_{s},L_{l}}Z(L_{b},L_{l}-L_{s})z^{L}\mu^{L_{l}-L_{s}}\label{eq:GC_Q1},\end{equation}
where $Z(L_{b},L_{s}-L_{l})$ is the canonical partition sum. Note
  that for $\mu=1$, Eq. (\ref{eq:GC_Q1}) is the grand canonical
  partition function of the Poland-Scheraga model extended to include
  all possible supercoil segment insertions. While this partition sum
  is different from that of the original PS model, it qualitatively
  yields the same phase diagram~\cite{KOM09}.

\begin{figure}
\begin{centering}
\includegraphics[scale=0.45]{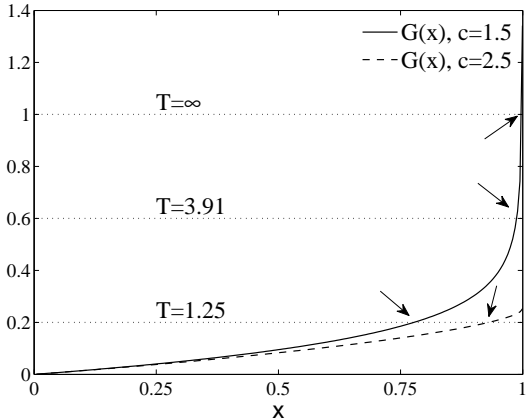} \par\end{centering}

\caption{\label{fig:transition}A plot of $G(x)$, as defined in (\ref{eq:Gx_def}),
as function of $x$. The solid line corresponds to $c=1.5$ while the
dashed line to $c=2.5$. The horizontal lines corresponds to different
temperatures. The
arrows point to the solutions of Eq.(\ref{eq:z_constraint3}) which
is equivalent to the thermodynamic limit. While
for $c=1.5$ there exists a solution for all temperatures, for $c=2.5$
there is a solution only up to some finite temperature, in this case
$\nu_{c}^{1/2}\omega_{c}^{-1}\approx0.25$. The parameters which were
used in this plot are $E_{b}=-3$, $E_{s}=-2$, $s=5$ and $A=0.1$.}

\end{figure}

The values of $z$ and $\mu$ are set by the conditions

\begin{eqnarray}
L & = & \frac{\partial \log Q}{\partial
  \log z}\label{eq:z_constraint}\ \ (=L_b+L_s+L_l),\\ 0 & =
& \frac{\partial \log Q}{\partial
  \log \mu}\label{eq:mu_constraint}\ \ (=L_l-L_s).\end{eqnarray}
Assuming that there is at least one bounded base pair, the grand partition sum can be written as

\begin{eqnarray}
Q(z,\mu) & = &
\tilde{V}(z,\mu)+\tilde{V}(z,\mu)U(z\mu)\tilde{V}(z,\mu)+...\\ & = &
\frac{\tilde{V}(z,\mu)}{1-\tilde{V}(z,\mu)U(z\mu)}\label{eq:GC_Q2},\end{eqnarray}
with

\begin{eqnarray}
\tilde{V}(z,\mu) & = &
\frac{V(z)}{1-V(z)W(z/\mu)}\label{eq:GC_Vt},\\ V(z) & = &
\sum_{n=1}^{\infty}\left(\omega z\right)^{n}=\frac{\omega z}{1-\omega
  z}\label{eq:GC_V}, \\
  W(z/\mu) & = & \sum_{n=1}^{\infty}\left(\nu
\frac{z}{\mu}\right)^{n}=\frac{\nu z}{\mu-\nu z}\label{eq:GC_W}, \\
U(z\mu) & = & \sum_{n=1}^{\infty}A\frac{\left(sz\mu\right)^{n}}{n^{c}}=A\Phi_{c}(sz\mu)\label{eq:GC_U}.\end{eqnarray}
The functions $U,V$ and $W$ represent the grand partition sums for
loops, bound segments and supercoils respectively. The polylog function $\Phi_{c}(q)$ is given by
\begin{equation}
\Phi_{c}(q)=\sum_{n=1}^{\infty}\frac{q^{n}}{n^{c}}.
\end{equation}
It is an analytic function everywhere except for a
branch cut at $q\in[1,\infty)$ \cite{Lewin1981}. It satisfies the
  relation
\begin{eqnarray}
\frac{d}{dq}\Phi_{c}(q) & = &
\frac{1}{q}\Phi_{c-1}(q)\label{eq:polylog_deriv}\ .
\end{eqnarray}
By inserting Eqs. (\ref{eq:GC_Vt}-\ref{eq:GC_U}) into
(\ref{eq:GC_Q2}), $Q(z,\mu)$ can be written as
\begin{equation}
Q(z,\mu)=\left[\left(\frac{1}{\omega z}-1\right)-\frac{\nu z}{\mu-\nu
    z}-A\Phi_{c}(sz\mu)\right]^{-1}\label{eq:GC_Q3}.\end{equation}
From this explicit form the constraints given by
Eqs.(\ref{eq:z_constraint}-\ref{eq:mu_constraint}) are readily
transformed into
\begin{eqnarray}
\left(\frac{1}{\omega z}-1\right)-\frac{\nu z}{\mu-\nu z} & = &
A\Phi_{c}(sz\mu)\label{eq:z_constraint2},\\ \frac{\nu z}{(\mu-\nu
  z)^{2}} & = &
\frac{A}{\mu}\Phi_{c-1}(sz\mu)\label{eq:mu_constraint2},\end{eqnarray}
where $z$ and $\mu$ from here on refer to the corresponding values in
the thermodynamic limit ($L\to\infty$) which is assumed in the
derivation of Eq.(\ref{eq:z_constraint2}).
Denoting by $m_{b}$, $m_{s}$ and $m_{l}$ the average density of
base pairs in bound segments, supercoils and loops, respectively, one
finds that
\begin{eqnarray}
m_{b} & = & -\frac{\partial \log z}{\partial \log \omega}\label{eq:mb_formula},\\
m_{s} & = & -\frac{\partial \log z}{\partial \log \nu}\label{eq:ms_formula},\\
m_{l} & = & -\frac{\partial \log z}{\partial \log s}\label{eq:ml_formula}.\end{eqnarray}
It is more convenient to work with the transformed variables $x=sz\mu$
and $y=\nu z/\mu$. Physically $\left(x/s\right)$ is the fugacity
associated with a unit increase in the total loop length, and
$\left(y/\nu\right)$ is the similar fugacity of supercoils. Under this
change of variables Eqs.(\ref{eq:z_constraint2},
\ref{eq:mu_constraint2}) become
\begin{eqnarray}
\frac{\sqrt{s\nu}}{\omega\sqrt{xy}}-\frac{1}{1-y} & = & A\Phi_{c}(x)\label{eq:z_constraint3},\\
\frac{y}{(1-y)^{2}} & = & A\Phi_{c-1}(x)\label{eq:mu_constraint3}.\end{eqnarray}
Considering $y$ as a function of $x$ through Eq.(\ref{eq:mu_constraint3}), let
\begin{equation}
G(x)\equiv\sqrt{\frac{xy}{s}}\left[A\Phi_{c}(x)+\frac{1}{1-y}\right]\label{eq:Gx_def},\end{equation}
so that Eq.(\ref{eq:z_constraint3}) can be written as
\begin{equation}
G(x)=\nu^{1/2}\omega^{-1}=e^{\beta\left(E_{b}-\frac{1}{2}E_{s}\right)}\equiv
H(T)\label{eq:z_constraint4}.\end{equation} Note that $y$ and $G(x)$
are increasing functions of $x$ in the physically relevant regime
$0\le x \le 1$.
The lower bound $x=0$ is achieved at zero temperature since
$\lim_{T\to 0}H(T) = 0$, while the upper bound is unity since
$\Phi_c(x)$ is a divergent sum for $x>1$. The presence of a
thermodynamic phase transition then depends on whether $x(T_c)=1$ is
achieved for some finite temperature $T_c$. Two regimes emerge as
shown in Fig.\ref{fig:transition}:\\ (i) $c\le2$: $\lim_{x\to
  1}\Phi_{c-1}(x)=\infty$, therefore
Eqs.(\ref{eq:z_constraint3},\ref{eq:mu_constraint3}) have a solution
in the interval $0\le x<1$ at all temperatures,\\ (ii) $c>2$: Note
that $\Phi_\alpha(1)=\sum_n 1/n^\alpha = \zeta_\alpha$ is the Riemann
Zeta function~\cite{AS1964}. Then, as $x\to 1$, the RHS of
Eqs.(\ref{eq:z_constraint3},\ref{eq:mu_constraint3}) remain finite
since $\infty>\zeta_{c-1}>\zeta_c>1$. For suitable values of $A$ and
$s$ (which guarantee $G(1)<1$), there exists a temperature $T_c$ such
that $x(T\ge T_c)=1$. The resulting nonanalyticity at $T_c$ translates
into the singular behavior of other quantities like the density
$m_{b}$ which underlies the melting transition. $T_{c}$ is
given by
\begin{equation}
G\left(1\right)=\nu_{c}^{1/2}\omega_{c}^{-1}=e^{\frac{1}{T_{c}}\left(E_{b}-\frac{1}{2}E_{s}\right)}\label{eq:Tc_eq}.\end{equation}
Below $T_{c}$, the system is fully defined by Eqs.(\ref{eq:z_constraint3}, \ref{eq:mu_constraint3}). Above $T_c$ one has
\begin{equation}
x = sz\mu=1\ ,\label{eq:z_mu_above_transition}\end{equation} and an
additional equation is necessary to impose the two constraints above.
To this end, we follow Ref.\cite{KMP2002} and introduce a
cutoff $M$ on the maximal loop size. In this reduced ensemble the
partition function is analytic, thus Eqs.(\ref{eq:z_constraint2},
\ref{eq:mu_constraint2}) are valid at all temperatures. The limit $M\to\infty$ reveals precisely how these equations are
modified above $T_c$, as discussed below.

\subsection{Regularizing the Grand Canonical Ensemble}

Introducing an upper cutoff $M$ on the allowed loop size, the loop
partition sum $U(z\mu)$ is replaced by
\[
U_{M}(z\mu)=A\sum_{n=1}^{M}\frac{(sz\mu)^{n}}{n^{c}}\equiv
A\Phi_{c}^{M}(sz\mu),
\]
where $\Phi_{c}^{M}(q)$ is the ``loop-truncated''
Polylog function, while the relation
$\frac{d}{dq}\Phi_{c}^{M}(sq)=\frac{1}{q}\Phi_{c-1}^{M}(sq)$ still
holds. The grand canonical partition sum is then
\[
Q_{M}(z,\mu)=\frac{\tilde{V}(z,\mu)}{1-\tilde{V}(z,\mu)U_{M}(z\mu)},
\]
and Eqs.(\ref{eq:z_constraint},\ref{eq:mu_constraint}) for the
constraints can be written as
\begin{eqnarray}
\left(\frac{1}{\omega z}-1\right)-\frac{\nu z}{\mu-\nu z} & = &
A\Phi_{c}^{M}(sz\mu)\label{eq:z_const_cutoff},\\ \frac{\nu z}{(\mu-\nu
  z)^{2}} & = &
\frac{A}{\mu}\Phi_{c-1}^{M}(sz\mu)\label{eq:mu_const_cutoff}.\end{eqnarray}
These equations hold for all $T$, since $\Phi_{c}^{M}(x)$ is an analytic
function. Our goal now is to analyze these equations in the limit $M\to\infty$ for temperatures above $T_{c}$. In this approach one should, in fact, consider the grand canonical ensemble with a finite but large average length of the DNA molecule $\left<L\right>$. One should then consider the limit $M,\left<L\right>\rightarrow \infty$ with $M\sim O(\left<L\right>)$. While considering finite $\left<L\right>$, Eq.(\ref{eq:z_const_cutoff}) is no longer exact but has $O\left(\left<L\right>^{-1}\right)$ correction. However, this correction does not modify the analysis presented below as it vanishes in the limit $M\sim \left<L\right> \rightarrow \infty$. We therefore take $\left<L\right>=\infty$ and then $M\rightarrow \infty$.

Let $T_c^M$ be the
temperature at which $sz\mu=1$ in this loop-truncated model, so that
for $T>T_c^M$ we have $sz\mu = 1 + \epsilon(M,T)$ with
$\epsilon>0$. Clearly,
as $M\to\infty$, $T_{c}^{M}\to T_{c}$ and $\epsilon\to0$ so that
$sz\mu=1$ for all $T>T_{c}$.  Then, for a given temperature
$T>sup_M\left(T_{c}^{M}\right)$ we have
\begin{eqnarray}
\frac{1}{\omega z}-\frac{\mu}{\mu-\nu z} & = &
A\left(\Phi_{c}^{M}(1)+b(\epsilon)\right)\label{eq:cutoff_eps_eq1},\\
\frac{\nu z}{\left(\mu-\nu z\right)^{2}} & = &
\frac{A}{\mu}\left(\Phi_{c-1}^{M}(1)+a(\epsilon)\right)\label{eq:cutoff_eps_eq2},
\end{eqnarray}
where
\begin{equation}
sz\mu  =  1+\epsilon\label{eq:cutoff_eps_eq3},\end{equation} and where
$a(\epsilon)$ and $b(\epsilon)$ are cutoff-dependent corrections
at least one of which is nonzero (otherwise the system is overdetermined).
We continue by assuming that $M\epsilon\to\infty$ as $M\to\infty$, and checking that this assumption is self consistent. With this assumption the leading behavior of $a(\epsilon)$ and $b(\epsilon)$ is found as
\begin{eqnarray}
a(\epsilon) & = &
\sum_{n=1}^{M}\frac{(1+\epsilon)^{n}-1}{n^{c-1}}\nonumber \\ & \approx
& \int^{M}dx\frac{e^{\epsilon x}-1}{x^{c-1}}\nonumber \\ & \sim &
\frac{e^{\epsilon
    M}}{M^{c-1}}\left(\frac{1}{\epsilon}+O\left(\frac{1}{M}\right)\right)\label{eq:a_eps},\end{eqnarray}
and
\begin{equation}
b(\epsilon)=\sum_{n=1}^{M}\frac{(1+\epsilon)^{n}-1}{n^{c}}\sim a(\epsilon)/M .
\end{equation}
Therefore,
the only cutoff-independent choice is
$a(0)=a_{0}$ and $b(0)=0$ for some constant $a_{0}$. Moreover, the
asymptotic form of $\epsilon$ as a function of $M$ follows from
Eq.(\ref{eq:a_eps}) as
\begin{eqnarray}
\epsilon(M,T) & = &
(c-2)\frac{\log M}{M}+O\left(\frac{\log \log M}{M}\right)\label{eq:eps_formula},\end{eqnarray}
demonstrating the self-consistency of the assumption above (see also
\cite{EH2005}).  We conclude that while for $T<T_{c}$
Eqs.(\ref{eq:z_constraint2}-\ref{eq:mu_constraint2}) hold, above $T_c$
they are replaced by
\begin{eqnarray}
\frac{1}{\omega z}-\frac{\mu}{\mu-\nu z} & = & A\zeta_{c},\\ \frac{\nu
  z}{\left(\mu-\nu z\right)^{2}} & = &
\frac{A}{\mu}\left(\zeta_{c-1}+a_{0}\right),\\ sz\mu & = &
1,\end{eqnarray} from which we can extract $a_{0}$

\begin{equation}
a_{0}=\frac{\nu}{As\left(\mu-\nu
  z\right)^{2}}-\zeta_{c-1}\label{eq:a0_formula}.\end{equation}
It measures the density of base pairs that
reside within a \emph{macroscopic loop }- or \emph{condensate }- that
appears above $T_c$. In the next section we discuss the order of the phase transition, where we take into account the condensate correction which was omitted in Ref. \cite{KOM09}

\subsection{Order of the Transition}

We show below that the above phase transition is continuous and then investigate the nature of the singularity at $T_c$. Consider the fraction of base pairs in bound segments, i.e., $m_{b}=-\partial log(z)/\partial log(\omega)$. Defining
\begin{equation}
P \equiv Q^{-1}(z,\mu;\omega)=\frac{1}{\omega z}-\frac{\mu}{\mu-\nu z}-A\Phi_{c}(sz\mu)\label{eq:Pdefine},
\end{equation}
 and noting that $P(z,\mu;\omega)=0$ for the poles of the
   partition function, the smallest of which yields the thermodynamic
   limit, we find
\begin{equation}
0=\frac{dP}{d\omega}=\frac{\partial P}{\partial\omega}+\frac{\partial
  P}{\partial z}\frac{\partial z}{\partial\omega}+\frac{\partial
  P}{\partial\mu}\frac{\partial\mu}{\partial\omega}.
\end{equation}
Rearranging and using Eq.(\ref{eq:mb_formula})
\begin{equation}
m_{b}=\frac{\omega}{z}\frac{\partial P/\partial\omega}{\partial
  P/\partial z + \left(\partial P/\partial \mu\right)\left(\partial \mu/\partial z\right)} \label{eq:formal_mb}.
\end{equation}
Evaluating the derivatives, making use of Eqs.(\ref{eq:z_constraint2},\ref{eq:mu_constraint2}, \ref{eq:z_mu_above_transition},\ref{eq:Pdefine}) we find that both below and above the critical temperature $m_b$ is given by
\begin{eqnarray}
m_{b}&=&\left[1+\frac{2\omega\nu\mu z^{2}}{\left(\mu-\nu z\right)^{2}}\right]^{-1}\label{eq:explicit_mb},\\
&\equiv&\left[1+\frac{2\omega y \sqrt{xy}}{\sqrt{s\nu}\left(1-y\right)^{2}}\right]^{-1}\label{eq:explicit_mb_in_x},
\end{eqnarray}
Below $T_c$ this can be obtained by noting that Eq.(\ref{eq:mu_constraint2}) implies $\frac{\partial P}{\partial\mu}=0$. Equation (\ref{eq:Pdefine}) can then be used to calculate $\frac{\partial P}{\partial\omega}$ and $\frac{\partial P}{\partial z}$ and finally Eq.(\ref{eq:mu_constraint2}) is used again to eliminate the polylog function. Above $T_c$ Eq.(\ref{eq:mu_constraint2}) does not hold and is replaced by $sz\mu=1$. Equation (\ref{eq:explicit_mb}) is then obtained by evaluating the partial derivatives appearing in Eq.(\ref{eq:formal_mb}). Equation (\ref{eq:explicit_mb}) implies that the order parameter $m_b$ is continuous across the transition since $z$ and $\mu$ are continuous functions of the temperature. Thus the transition is continuous.

For a detailed analysis of the singularity it is convenient to express $m_b$ in terms of the $x,y$ variables. Let $\delta x$,
$\delta y$, and $\delta m_b$ denote the deviation of $x$, $y$, and
$m_b$, respectively, from their values at $T_c$ due to a slight change
in temperature $t=T-T_c$. First we explore the relations among $t$, $\delta x$ and $\delta y$. Above the transition $x=1$ and hence $\delta x=0$. Thus $\Phi_c(x)$ in Eq.(\ref{eq:z_constraint3}) becomes $\zeta_c$. As a result $y$ has a power series expansion above $T_c$ where to leading order $\delta y \propto t$. Hence for $t>0$, $\delta m_b=f(t)$ where $f$ is analytic near $t=0$. Below the transition one has to make use of the expansion of the polylog function
\begin{equation}
\Phi_{c-1}(1-\delta x)=\zeta_{c-1}+\zeta_{c-2}\delta x+...+\Gamma(c-2)\delta x^{c-2} +...
 \label{eq:polylog_expension}\end{equation}
 where $\Gamma(c-2)$ is the Gamma function and the last term is the leading singular term in the expansion. We proceed by separately considering two regimes of the parameter $c$.

\begin{itemize}
\item For $2<c<3$, the expansion (\ref{eq:polylog_expension}) becomes $\Phi_{c-1}(1-\delta x)\approx \zeta_{c-1}+\Gamma(c-2)\delta x^{c-2}$ and therefore Eq.(\ref{eq:mu_constraint3}) yields
    \begin{equation}\delta y \sim \delta x^{c-2}\label{eq:deltay_deltax_relation},\end{equation}
    which implies $\delta x \ll \delta y$. Thus in the vicinity of the transition temperature Eqs.(\ref{eq:z_constraint3}, \ref{eq:explicit_mb_in_x}) yield $\delta m_b \approx f(t)+ \tilde{\alpha} \delta x $ where $f(t)$ is the same function as above the transition, and $\tilde{\alpha}$ is a constant. Using (\ref{eq:deltay_deltax_relation}) and noting that $\delta y \propto t$ one finally obtains $\delta m_b = f(t) + \alpha t^{\frac{1}{c-2}}$ where $\alpha$ is a constant. Since $f(t)$ is analytic function, the $\left\lceil \frac{1}{c-2}\right\rceil$ derivative of $m_b$ is discontinuous and the transition is of order $\left\lceil\frac{c-1}{c-2}\right\rceil$ .\\
\item For $c\ge 3$, the expansion of the polylog function is $\Phi_{c-1}(1-\delta x)\approx \zeta_{c-1}+\zeta_{c-2}\delta x$. Hence Eq.(\ref{eq:mu_constraint3}) yields $\delta y \propto \delta x$, which together with (\ref{eq:z_constraint3}) implies $\delta x \propto \delta y \propto t$. Thus $\delta m_b \approx f(t)+\gamma t$ where $\gamma$ is a constant. This implies that the first derivative of  $\delta m_b$ is discontinuous, and the transition is of second order.\\
\end{itemize}

In summary, the transition is characterized by the singular behavior of $\delta m_b$ below:
\begin{equation}
\delta m_b  =  \left\{
\begin{array}{lcc} f(t) & \qquad &  t>0 \\
f(t) + \alpha t^\eta & & t>0\\ \end{array}
\right.,
\end{equation}
with
\begin{equation}
\eta = \left\{
\begin{array}{ccr} \frac{1}{c-2} & & 2<c<3 \\
1 & \qquad & c\ge 3\\ \end{array}
\right.,
\end{equation}
where $f(t)$ can be expressed as a power series in $t$ for $t>0$. Since $m_l=m_s=(1-m_b)/2$, a similar singular behavior is exhibited by these variables. Hence the denaturation transition of a circular DNA is
second order for $c\ge 3$, third order for $2.5\le c < 3$,
forth order for $\frac{7}{3}\le c<2.5$, etc., approaching infinite order as $c\to 2$. No phase transition takes place  for $c\le 2$. In contrast, a DNA without helicity (as described by the original
PS model) melts through a first order transition for $c>2$ and a second order transition for $1<c\le 2$.

\subsection{High Temperature Phase}

The high-temperature phase of the PS model is composed of an
all-encompassing macroscopic loop created at $T_c$ through a jump in
the loop fraction to its maximum value $m_l=1$. Here, we not only have
a smoother transition but also a qualitatively different denatured
phase. For example, the loop fraction reaches its maximum value
($m_l=1/2$ within the present model) only as $T\to\infty$ and it
continuously increases across and above $T_c$. At this point, one is
tempted to ask what has changed qualitatively across the
transition. In this section, we show that a macroscopic loop is again
the distinguishing feature.  However, instead of being an all-or-none
phenomenon, the dominance of the macro-loop among the denatured base
pairs grows steadily from $T_c$ on. Below we analyze the loop length
distribution, $p_M(l)$, demonstrating that in addition to the power
law behavior on microscopic scale, it exhibits a peak at lengths of
order $M$ whose integrated weight is of order $1/M$. This peak
represents the macroscopic loop which opens up above $T_c$. Note
  that the probability distribution functions for bound and
  supercoiled segment lengths are still exponential in the length $n$,
  since the corresponding Boltzmann weights are $(\omega z)^n$ and
  $y^n$, respectively.

The loop size distribution in the ``loop-truncated'' model is given
by
\begin{equation}
p_{M}(l)=\frac{1}{\Phi_{c}^{M}(sz\mu)}\frac{(sz\mu)^{l}}{l^{c}}\Theta(M-l)\label{eq:loop_prob_cutoff},\end{equation}
where $\Theta(x)$ is the Heaviside function. Differentiating with
respect to $l$ and noting that $sz\mu=e^{\epsilon}$ we find that
this distribution exhibits a minimum at
\begin{equation}
    l^*=\frac{c}{\epsilon}=\frac{c}{c-2}\frac{M}{\log M},
\end{equation}
where Eq.(\ref{eq:eps_formula}) has been used. The large $l$
distribution is peaked at $l=M$ with $p_M(M)\approx
\zeta_c^{-1}e^{\epsilon M}/M^c\sim M^{-2}$. Thus the integrated
weight of the peak is $O(1/M)$ up to logarithmic corrections.
Since as discussed above $M\sim O(L)$, One expects $O(1)$ number of macroscopic loops to open up above
$T_c$. In fact one can argue that entropy favors a single
macroscopic loop \cite{EH2005}. To see this one can compare the probability of a state with only one macroscopic loop with that of configurations with two macroscopic loops: Assuming that there are $l_{con}\sim M$ base pairs within the condensed phase, the weight of configurations with single loop is
\begin{equation*}
\Gamma_1\left(l_{con}\right)\approx L_l p_M\left(l_{con}\right)\sim O\left(L^{1-c}\right)(sz\mu)^{l_{con}}.
\end{equation*}
The weight of configurations with two macroscopic loops is
\begin{eqnarray*}
\Gamma_2\left(l_{con}\right) & \approx & \binom {L_l}{2} \sum_{n\sim M} p_M\left(n\right)p_M\left(l_{con}-n\right) \\
& \sim & O\left(L^{3-2c}\right)(sz\mu)^{l_{con}}.
\end{eqnarray*}
As $c>2$, it follows that configurations with a single macroscopic loop dominates the ensemble in the limit $M\sim L\rightarrow \infty$.

\begin{figure}
\begin{centering}
\includegraphics[scale=0.5]{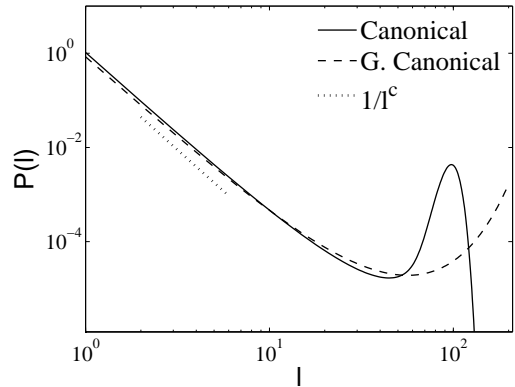} \par\end{centering}

\caption{\label{fig:Pl}The loop size distribution $p(l)$ in the canonical (solid
  line) and the regularized grand-canonical (dashed line)
  ensembles. For small values of $l$ the critical phase can be
  identified (where $p(l)\sim l^{-c}$). For $l\sim L$ the canonical
  curve shows the {}``bump'' around $l\approx\xi$ (see text) while the
  grand-canonical curve behaves in a somewhat different manner. The
  parameters which were used for this plot are $c=3.5$, $L=400$, $M=200$,
  $E_{b}=-3$, $E_{s}=-2$, $s=5$ and $A=0.1$ and $T=3$ ($T_{c}=1.167$)}

\end{figure}

The condensation phenomenon observed in this model is reminiscent of
condensation in Bose-Einstein Gas and the zero-range process (ZRP)
when the density of particles is above a critical value
~\cite{EH2005,EMZ2006}. Figure \ref{fig:Pl} shows the
loop size distribution for finite $M$ and $T>T_{c}$. A power-law decay
with the exponent $c$ for $l\ll M$ and a peak for $l\lesssim M$ which
is the precursor of the $\delta$-function representing the macroscopic
loop are evident.

\section{Canonical Treatment}
\label{sec:canonical}
\begin{figure*}
\begin{centering}
\includegraphics[scale=0.5]{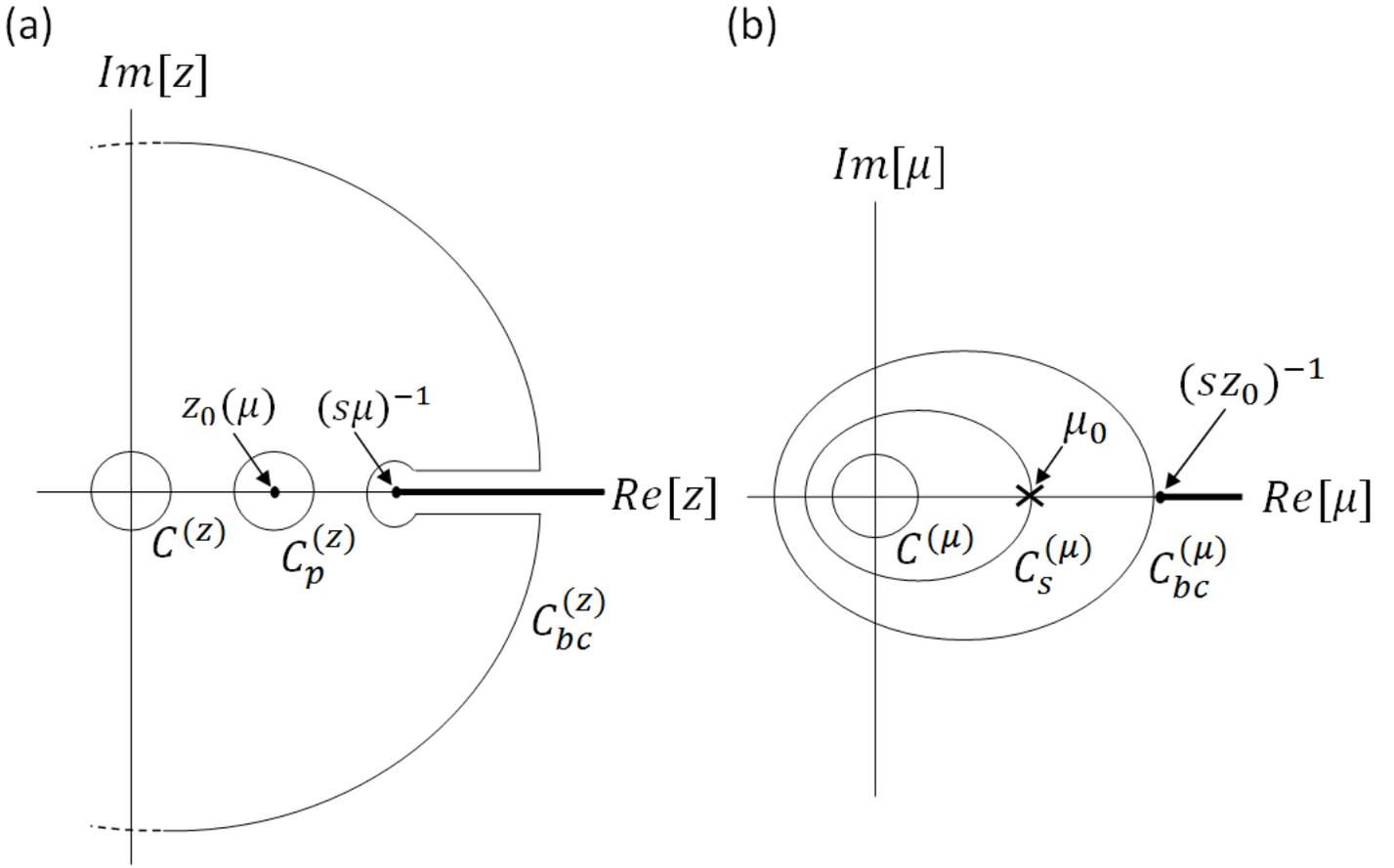}
\par\end{centering}

\caption{\label{fig:contour} The integration procedure used for the
  canonical partition function: (a) In the $z$ plane, the contour
  $C^{(z)}$ which encircles the origin can be replaced by
  $C_{p}^{(z)}$ around the pole at $z_0$ and $C_{bc}^{(z)}$ which
  wraps the branch cut (thick line) and closes at infinity; (b) in the $\mu$ plane, the contour $C^{(\mu)}$
  can be deformed to pass through a saddle point $\mu_0$ when it
  exits. Otherwise the dominant contribution comes from the vicinity of
  the branch point (see text).}

\end{figure*}

In order to justify the regularization procedure applied in the
grand canonical ensemble we study the model within the canonical
ensemble, namely with fixed $L$ and $L_{l}-L_{s}=0$. In addition
this approach allows us to study the properties of the condensate
and to further illuminate the mathematical structure underlying the
phase transition. The canonical partition function can be obtained
from the grand sum in Eq.(\ref{eq:GC_Q1}) by means of Cauchy
integration:
\begin{equation}
Z(L,L_{l}-L_{s})=\frac{1}{\left(2\pi
  i\right)^{2}}\oint_{C^{(\mu)}}d\mu\oint_{C^{(z)}}dz\frac{Q\left(z,\mu\right)}{z^{L+1}\mu^{L_{l}-L_{s}+1}}\label{eq:can_Z_int},\end{equation}
where $C^{(\mu)}$ and $C^{(z)}$ are circular, counter-clockwise
oriented contours which are centered at the origin and enclose no
singularity of $Q(z,\mu)$ (Fig.\ref{fig:contour}).  Enforcing the
linking number constraint, $L_{l}=L_{s}$, and using
Eq.(\ref{eq:GC_Q3}) yield
\begin{eqnarray}
Z(L,0) & = & \frac{1}{\left(2\pi i\right)^{2}}\oint d\mu\oint dz\ I(z,\mu)\label{eq:can_Z},\\
I(z,\mu) & = & \frac{\left[\frac{1}{\omega z}-1-\frac{\nu z}{\mu-\nu z}-A\Phi_{c}(sz\mu)\right]^{-1}}{z^{L+1}\mu}\label{eq:I_def}.\end{eqnarray}
For $|\mu|$ sufficiently small so that $|sz\mu|<1$, let $z_0$ be the
nontrivial pole of $I(z,\mu)$ in the $z$-plane, given by
Eq.(\ref{eq:z_constraint2}). Then, by Cauchy's integral theorem, the
integration contour $C^{(z)}$ can be replaced by
$C_{p}^{(z)}+C_{bc}^{(z)}$ shown in Fig.\ref{fig:contour}. Due to the
factor $z^{-L}$ in (\ref{eq:I_def}) the dominant contribution comes
from $C_{p}^{(z)}$ and we obtain
\begin{equation}
Z(L,0)\sim\frac{1}{2\pi i}\oint_{C^{(\mu)}}\frac{d\mu}{\mu}z_0(\mu)^{-L-1}\label{eq:can_Z_mu_int}.
\end{equation}

We now evaluate the integral separately below and above the critical point. Below the transition,
the integrand in Eq.(\ref{eq:can_Z_mu_int}) has a saddle point
given by $\frac{dz_0}{d\mu}|_{\mu_0}=0$ and
$|\mu_0|<1/s|z_0|$. The partition function can now be evaluated by
first deforming $C^{(\mu)}$ into the contour $C_{s}^{(\mu)}$ which
passes through this saddle point (Fig.\ref{fig:contour}) and then
approximating the integral by the contribution from the vicinity of
$\mu_0$, i.e.,
\begin{equation}
Z(L,0)\sim
e^{-L\log z_0(\mu_0)}\label{eq:Z_can_solution}.
\end{equation}
After differentiating Eq.(\ref{eq:z_constraint2}) with respect to
$\mu$ and setting $\frac{dz_0}{d\mu}=0$ we find
Eq.(\ref{eq:mu_constraint2}) as the saddle-point condition. These two
equations fix $z_0$ and $\mu_0$ and describe the system for $T<T_c$,
as was found earlier in the grand-canonical framework. Note that the
free energy is obtained from $z_0$ through
Eq.(\ref{eq:Z_can_solution}).

Above the critical temperature, this procedure is not
applicable,
as the solution of $\frac{dz_0}{d\mu}=0$ for $\mu_0$ now lies on the
branch cut. However, it is found that Eq.(\ref{eq:Z_can_solution})
holds, with $z_0$ and $\mu_0$ given now by
Eqs.(\ref{eq:z_constraint2},\ref{eq:z_mu_above_transition}) rather
than (\ref{eq:z_constraint2},\ref{eq:mu_constraint2}) as obtained
within the grand canonical ensemble. This can be shown by evaluating
the integral in Eq.(\ref{eq:can_Z_mu_int}) along another contour
$C_{bc}^{(\mu)}$ shown in Fig.\ref{fig:contour} on which
$|sz_0\mu|\lesssim 1$. After a change of variables
$e^{-u}=sz_0\mu$, Eq.(\ref{eq:can_Z_mu_int}) transforms to
\begin{equation}
Z(L,0)\sim \oint
\frac{du}{2\pi i}\ \alpha(u)\,e^{-L\log z_0(u)}\label{eq:can_Z_u_int},
\end{equation}
where $\alpha(u)=\frac{-1}{z_0}\left(1+\frac{d\log z_0}{du}\right)$
is a nonextensive correction to the free energy that can be neglected.
The main contribution along the contour $C_{bc}^{(\mu)}$ is from the
neighborhood of the branch cut where $|u|\ll1$ with $\mbox{Re}[u]$
positive and as small as desired. We therefore express $\log\,z_0(u)$
in terms of the small parameter $u$ by using the implicit equation
(\ref{eq:z_constraint2}) and the nonanalytic expansion of
$\Phi_{c}(1-u)$ given by Eq.(\ref{eq:polylog_expension}), to obtain
\[
\log z_0(u) \approx\sum_{n=0}^{\left\lfloor
  c-1\right\rfloor }b_{n}u^{n}+b_{c}u^{c-1}+...
\]
where ${b_n}$ are temperature dependent coefficients with
$b_{0}=\log z_0(0)$ and
$b_{n}=\frac{1}{n!}\frac{d^{n}\log z_0}{du^{n}}|_{u=0}$.
The coefficient of the linear term $b_1$ vanishes at $T_c$. This
follows directly from Eqs.(\ref{eq:z_constraint2},\ref{eq:mu_constraint2},\ref{eq:z_mu_above_transition}). It changes
sign from $b_1>0$ below the transition, where $sz_0\mu_0<1$, to
$b_1<0$ above it. Let $g(u)$ be the nonlinear part of the expansion
$g(u)\approx\sum_{n=2}^{\left\lfloor c-1\right\rfloor
}b_{n}u^{n}+b_{c}u^{c-1}$. Note that for $c<3$, $g(u)=b_{c}u^{c-1}$.
One therefore has
\begin{equation}
Z(L,0)\sim e^{-L \log z_0(0)}\oint
\frac{du}{2\pi i}\,e^{-L\left[b_{1}u+g(u)\right]}\label{eq:can_Z_u_int2}\ .
\end{equation}
As $u$ is approximately imaginary in the region of interest, the integrand is oscillatory, yielding vanishing contribution at large $L$ except in the small region where $\mbox{Im}[u]\lesssim
O\left(\frac{1}{L}\right)$. As a result one may expand the integrand in Eq.(\ref{eq:can_Z_u_int2}) as $e^{-L\left[b_{1}u+g(u)\right]}\approx e^{-b_{1}Lu}\left[1-Lg(u)\right]$. Moreover, the integration contour can be replaced by the right vertical tangent of $C_{bc}^{(\mu)}$ in Fig.\ref{fig:contour}.
Combining these observations we get
\begin{equation}
Z(L,0)\sim
e^{-L\log z_0(0)}\int_{-i\infty}^{i\infty}du
  \,e^{-b_{1}Lu}\left[1-Lg(u)\right] \label{eq:can_Z_expansion}.
\end{equation}

The analytic terms of the integrand do not contribute, since the
integration yields a delta function $\delta\left(b_{1}L\right)$ or
its derivatives~\cite{EMZ2006}. Therefore, the partition function is
determined solely by the nonanalytic term in $g(u)$ as:
\begin{eqnarray}
Z(L,0) & \sim &
e^{-L\log z_0(0)}b_{c}L\int_{-i\infty}^{i\infty} du\,
e^{-b_{1}Lu}u^{c-1}\nonumber \\
& = &
e^{-L\log z_0(0)}\frac{\tilde{b}}{b_{1}^{c}L^{c-1}}\label{eq:Z_can_above},\end{eqnarray}
where $\tilde{b}\equiv b_{c}\frac{sin\left(\pi
  c\right)}{\pi}\Gamma(c)$ \cite{EMZ2006}. The free energy density is,
of course, continuous across $T_c$ and above the critical temperature
it is determined by
Eqs.(\ref{eq:z_constraint2}) and (\ref{eq:z_mu_above_transition}),
as in the grand canonical treatment.
\subsection{High Temperature Phase}
In this subsection we consider the loop size distribution $p(l)$ at
temperatures above $T_c$. As in the grand canonical ensemble, a
condensate phase composed of a macroscopic loop is found, although
the details of the peak in $p(l)$ corresponding to this phase are
different. The analysis follows the analysis carried out for
the condensation transition in the zero-range process
\cite{EMZ2006}. Here we just outline the main results.

Within the canonical ensemble the loop size distribution is given by
\begin{equation}
  p(l)=A\frac{s^{l}}{l^{c}}\frac{Z(L-l,-l)}{Z(L,0)},
\end{equation}
where $Z(L-l,-l)$ is given by
\begin{eqnarray}
Z(L-l,-l)&=&\frac{1}{\left(2\pi i\right)^{2}}\oint_{C^{(\mu)}}d\mu\oint_{C^{(z)}}dz\frac{Q\left(z,\mu\right)}{z^{L+1-l}\mu^{1-l}} \nonumber \\
&\approx&\frac{s^{-l}}{2\pi i}\oint_{C^{(\mu)}}d\mu\left(sz_0\mu\right)^{l}e^{-L\log z_0(\mu)} \nonumber \\
&\approx&\frac{s^{-l}}{2\pi i}\oint_{C_{bc}^{(\mu)}}du\,e^{-L\left[\log z_0(u) +\phi u\right]} , \nonumber
\end{eqnarray}
with $\phi=l/L$ and $e^{-u} = sz_0\mu$. Expanding for small $u$ yields
\begin{eqnarray}
p(l) &\approx & \frac{A}{l^c}\frac{I(l/L)}{I(0)}, \label{eq:pl_form} \\
I(\phi) & \equiv & \frac{1}{2\pi i}\int_{-i\infty}^{i\infty}du\,e^{-L\left[(b_{1}+\phi)u+g(u)\right]} \label{eq:def_Iphi}.
\end{eqnarray}
For $T>T_c$, the function $I(\phi)$ develops a peak at $\phi\simeq
-b_1 \equiv \xi$. This is demonstrated separately for $2<c<3$ and $c
> 3$.

\textbf{(a)} For $2<c<3$, the leading-order term in $g(u)$ is the
nonanalytic term $u^{c-1}$ and $I(\phi)$ can be written in the form
\begin{eqnarray}
I(\phi)=L^{-1/(c-1)}V_c\left[L^{\frac{c-2}{c-1}}(\phi-\xi)\right]
\label{eq:Iphi_c2to3}.
\end{eqnarray}
The asymptotic behavior of the scaling function $V_c[q]$ are given by
\begin{eqnarray}
    V_c[q] \simeq \begin{cases}
            a |q|^{-c} & q\rightarrow -\infty \\
            c_1 q^{(3-c)/2(c-2)}e^{-c_2q^{(c-1)/(c-2)}} & q\rightarrow \infty,
            \end{cases} \label{eq:scalingV}
\end{eqnarray}
where the constants $a$, $c_1$ and $c_2$ are given in Eqs.(81-83) of
\cite{EMZ2006}. Equations(\ref{eq:pl_form}-\ref{eq:scalingV}) together with Eq.(\ref{eq:Z_can_above}) yield after some algebra \cite{EMZ2006}

\[
p(l)\sim\begin{cases}
\frac{l^{-c}}{(\xi - l/L)^{c}}   &   \xi L-l\gg O\left(L^{\frac{1}{c-1}}\right)\\
\frac{l^{-c}}{\left(\frac{l}{L}-\xi\right)^{\frac{c-3}{2(c-2)}}} e^{-c_2\left(\frac{l}{L}-\xi\right)^{\frac{c-1}{c-2}}}
& \ \ l-\xi L\gg O\left(L^{\frac{1}{c-1}}\right).\end{cases}
\]
In the intermediate regime where $|l - \xi L|\ll L^{\frac{1}{c-1}}$,
$p(l)$ has the form
\begin{equation} p(l)\sim
L^{-c/(c-1)}V_c\left[\frac{l-L\xi}{L^{1/(c-1)}}\right]
\label{eq:pl_condensate}.
\end{equation}
Therefore $p(l)$ has a peak centered around
$l\simeq \xi L$ with a power-law decay on the right and a stretched
exponential decay on the left. Integrating $p(l)$ as given by (\ref{eq:pl_condensate}) around $l\approx
\xi L$ yields an order $1/L$ contribution
implying the existence of a macroscopic loop.

\textbf{(b)} For $c > 3$ the resulting behavior is summarized in
Eqs.(100-101) of \cite{EMZ2006} and read
\[
    I(\phi)\simeq \begin{cases}
        \frac{a}{(\xi-\phi)^c L^{c-1}}  & \xi-\phi \sim O(1) \\
        \frac{1}{\sqrt{4\pi |b_2|L}}e^{L\frac{(\phi-\xi)^2}{4b_2}} & |\xi-\phi| \ll O\left(L^{-1/3}\right).
        \end{cases}
\]
Note that $b_2<0$. Hence $p(l)$ is of the form
\[
p(l)\sim\begin{cases}
\frac{l^{-c}}{(\xi-l/L)^{c}} & \xi L-l \sim O(L) \\
\frac{(l/L)^{-c}}{L\sqrt{L}}e^{\frac{(l-\xi L)^2}{4b_2 L}}
 & |\xi L -l| \ll O\left(L^{2/3}\right). \end{cases}
\]
Therefore in this case the condensate bump has a Gaussian form with
weight $O(1/L)$, as in the case $2<c<3$.

We conclude that the loop size distribution $p(l)$ is a power law
(reminiscent of the critical phase) for $O(1)$ loops, superposed
with a bump centered around $\xi L=|b_1| L$ as shown in
Fig.\ref{fig:Pl}. The precise form of this condensate peak differs
from the one found in the grand-canonical analysis, although both ensembles yield the same phase diagram in the large $L$ limit. This result is
very similar to what is found in the context of ZRP, however it is
not exactly the same. Within the ZRP, above the critical density any
further increase in the density is absorbed by the condensate. Here,
on the other hand, the total length of the loops in the critical phase changes with temperature above $T_c$. In particular, it is finite at $T_c$ and approaches zero at $T\rightarrow\infty$.
Since the loop size distribution in the critical phase is fixed above the critical
temperature, it implies that the number of loops in the critical phase varies with $T$.
\section{Conclusions}
\label{sec:conclusions}
We analyzed the denaturation transition of circular DNA chains,
assuming that opening denatured loops induces formation of
supercoils. As in the case of non-circular DNA the thermodynamic
behavior of the model is found to be determined by the loop entropy
parameter $c$. We find that for $c\le2$ the model exhibits no
transition while for $c>2$ the transition is continuous, of order
$\left\lceil\frac{c-1}{c-2}\right\rceil$. Thus for $c\ge 3$ the
transition is second order, while for $2<c<3$ (which includes the
physical value of $c\approx2.12$) it is of higher order reaching
$\infty$-order as $c\to 2$.

In addition, the nature of the denaturated phase is rather different
from that of the non-circular DNA. Here a macroscopic loop
(condensate) is formed above $T_c$ whose length increases continuously
as the temperature is increased. This is different from the
denaturated phase in the non-circular case, where the two strands are
fully separated at all temperatures above $T_c$. This is reminiscent
of Bose-Einstein condensation and to similar real space condensation
encountered in models such as the
ZRP~\cite{EH2005,EMZ2006}. Furthermore, the difference observed in the
condensate peaks of canonical and grand-canonical ensembles (for
finite $L$) has the same mathematical structure as in the ZRP.

A different mechanism for absorbing the extra linking number
produced by opening of loops in circular DNA has been considered
previously ~\cite{RB2002,GOY2004}. In this mechanism the extra
linking number is compensated by overtwist of remaining bound
segments of the molecule at the cost of an elastic energy. This
mechanism also yields smoothening of the denaturation transition as
obtained in the present paper. It would be of interest to consider
the denaturation transition in the case where both overtwist and
supercoils are present.

Finally, our results apply to a homogeneous polymer where there is a
single binding energy. It is well known that introducing disorder also
smoothens the first-order transition in the PS model
\cite{CY2007}. The influence of sequence inhomogeneity on
the present melting transition which is already smoothened by
topological constraints is an open question.

We thank O. Cohen, M.R. Evans, O. Hirschberg, S.N. Majumdar and E. Orlandini for helpful discussions. This work was supported by the Israel Science Foundation (ISF) and the Turkish Technological and Scientific Research
Council (TUBITAK) through the grant TBAG-110T618.


\begin{thebibliography}{10}

\bibitem{DSD1992}
D.~Dixon, R.~Simpson-White, and L.~Dixon,
\newblock J. Mar. Biol. Assoc. UK {\bf 72}, 519 (1992).

\bibitem{HS2004}
D.~Hickey and G.~Singer,
\newblock Genome biology {\bf 5}, 117 (2004).

\bibitem{WB1985}
R.~M. Wartell and A.~S. Benight,
\newblock Physics Reports {\bf 126}, 67 (1985).

\bibitem{HM1994}
H.~Hiasa and K.~Marians,
\newblock Journal of Biological Chemistry {\bf 269}, 32655 (1994).

\bibitem{HM1996}
H.~Hiasa and K.~Marians,
\newblock Journal of Biological Chemistry {\bf 271}, 21529 (1996).

\bibitem{CH2006}
E.~Carlon and T.~Heim,
\newblock Physica A: Statistical Mechanics and its Applications {\bf 362}, 433
  (2006).

\bibitem{PB1989}
M.~Peyrard and A.~R. Bishop,
\newblock Phys. Rev. Lett. {\bf 62}, 2755 (1989).

\bibitem{Fisher1966}
M.~E. Fisher,
\newblock J. Chem. Phys. {\bf 45}, 1469 (1966).

\bibitem{PS1966}
D.~Poland and H.~A. Scheraga,
\newblock J. Chem. Phys. {\bf 45}, 1456 (1966).

\bibitem{DPB1993}
T.~Dauxois, M.~Peyrard, and A.~R. Bishop,
\newblock Phys. Rev. E {\bf 47}, 684 (1993).

\bibitem{KMP2000}
Y.~Kafri, D.~Mukamel, and L.~Peliti,
\newblock Phys. Rev. Lett. {\bf 85}, 4988 (2000).

\bibitem{WM1972}
R.~Wartell and E.~Montroll,
\newblock Adv. Chem. Phys. {\bf 22}, 129 (1972).

\bibitem{Ben1980}
C.~J. Benham,
\newblock J. Chem. Phys. {\bf 72}, 3633 (1980).

\bibitem{Ben1992}
C.~J. Benham,
\newblock Journal of Molecular Biology {\bf 225}, 835  (1992).

\bibitem{BM2000}
C.~Bouchiat and M.~Mezard,
\newblock The European Physical Journal E: Soft Matter and Biological Physics
  {\bf 2}, 377 (2000).

\bibitem{PMD2007}
J.~Palmeri, M.~Manghi, and N.~Destainville,
\newblock Phys. Rev. Lett. {\bf 99}, 088103 (2007).

\bibitem{MS1994b}
J.~F. Marko and E.~D. Siggia,
\newblock Macromolecules {\bf 27}, 981 (1994).

\bibitem{NG2006}
R.~A. Neher and U.~Gerland,
\newblock Phys. Rev. E {\bf 73}, 030902 (2006).

\bibitem{KMP2002}
Y.~Kafri, D.~Mukamel, and L.~Peliti,
\newblock Euro. Phys. J. B {\bf 27}, 135 (2002).

\bibitem{COS2002}
E.~Carlon, E.~Orlandini, and A.~L. Stella,
\newblock Phys. Rev. Lett. {\bf 88}, 198101 (2002).

\bibitem{Meltsim}
R.~Blake {\em et~al.},
\newblock Bioinformatics {\bf 15(5)}, 370 (1999).

\bibitem{RB2002}
J.~Rudnick and R.~Bruinsma,
\newblock Phys. Rev. E. {\bf 65}, 030902(R) (2002).

\bibitem{GOY2004}
T.~Garel, H.~Orland, and E.~Yeramian,
\newblock Arxiv preprint q-bio/0407036  (2004).

\bibitem{KOM09}
A.~Kabak\c{c}{\i}o\u{g}lu, E.~Orlandini, and D.~Mukamel,
\newblock Phys Rev E. {\bf 80}, 010903(R) (2009).

\bibitem{KOM2010}
A.~Kabak\c{c}{\i}o\u{g}lu, E.~Orlandini, and D.~Mukamel,
\newblock Physica A: Statistical Mechanics and its Applications {\bf 389}, 3002
   (2010).

\bibitem{KAS2010}
M.~Sayar, B.~Av{\c{s}}aro{\u{g}}lu, and A.~Kabak{\c{c}}{\i}o{\u{g}}lu,
\newblock Physical Review E {\bf 81}, 041916 (2010).

\bibitem{YI2002}
L.~Yan and H.~Iwasaki,
\newblock Japanese Journal of Applied Physics {\bf 41}, 7556 (2002).

\bibitem{Dup1986}
B.~Duplantier,
\newblock Phys. Rev. Lett. {\bf 57}, 941 (1986).

\bibitem{Lewin1981}
L.~Lewin,
\newblock {\em Polylogarithms and Associated Functions} (North-Holland
  Publishing Co., New York, 1981).

\bibitem{AS1964}
M.~Abramowitz and I.~Stegun,
\newblock {\em Handbook of Mathematical Functions}, Fifth ed. (Dover, New York,
  1964).

\bibitem{EH2005}
M.~Evans and T.~Hanney,
\newblock Journal of Physics A: Mathematical and General {\bf 38}, R195 (2005).

\bibitem{EMZ2006}
M.~R. Evans, S.~N. Majumdar, and R.~K.~P. Zia,
\newblock J. Stat. Phys. {\bf 123(2)}, 357 (2006).

\bibitem{CY2007}
B.~Coluzzi and E.~Yeramian,
\newblock The European Physical Journal B - Condensed Matter and Complex
  Systems {\bf 56}, 349 (2007).

\end{thebibliography}
\end{document}